\documentclass[11pt]{article}
\usepackage{geometry}
\usepackage{amsthm,amsmath,amsfonts,amssymb,bm, cases}
\usepackage{graphicx,psfrag,epsf}
\usepackage{caption}
\usepackage{tabularx,multicol,multirow}
\usepackage{enumerate}
\usepackage{natbib}
\usepackage{enumitem}
\usepackage{setspace}
\usepackage[affil-it]{authblk}
\usepackage{booktabs}

\def\I{\mathbb{I}}
\def\ci{\perp\!\!\!\perp}

\providecommand{\keywords}[1]
{
 \small	
 \textbf{\textit{Keywords:}} #1
}

\title{Bayesian Fractional Polynomial Approach to Quantile Regression and Variable Selection with Application in the Analysis of Blood Pressure among US Adults}

\author[1]{Sanna Soomro}
\author[1]{Keming Yu \thanks{keming.yu@brunel.ac.uk}}
\affil[1]{Brunel University London}

\date{}

\begin{document}
\maketitle

\begin{abstract}
Hypertension is a highly prevalent chronic medical condition and a strong risk factor for cardiovascular disease (CVD), as it accounts for more than $45\%$ of CVD. The relation between blood pressure (BP) and its risk factors cannot be explored clearly by standard linear models. Although the fractional polynomials (FPs) can act as a concise and accurate formula for examining smooth relationships between response and predictors, modelling conditional mean functions observes the partial view of a distribution of response variable, as the distributions of many response variables such as BP measures are typically skew. Then modelling ‘average’ BP may link to CVD but extremely high BP could explore CVD insight deeply and precisely. So, existing mean-based FP approaches for modelling the relationship between factors and BP cannot answer key questions in need. Conditional quantile functions with FPs provide a comprehensive relationship between the response variable and its predictors, such as median and extremely high BP measures that may be often required in practical data analysis generally. To the best of our knowledge, this is new in the literature. Therefore, in this paper, we employ Bayesian variable selection with quantile-dependent prior for the FP model to propose a Bayesian variable selection with parametric nonlinear quantile regression model. The objective is to examine a nonlinear relationship between BP measures and their risk factors across median and upper quantile levels using data extracted from the 2007-2008 National Health and Nutrition Examination Survey (NHANES). The variable selection in the model analysis identified that the nonlinear terms of continuous variables (body mass index, age), and categorical variables (ethnicity, gender and marital status) were selected as important predictors in the model across all quantile levels.
\end{abstract}


\setstretch{1.5} 

\section{Introduction}

Over the past three decades, the number of adults aged 30-79 with hypertension has increased from 648 million to 1.278 billion globally (\cite{ZhouEtAl2021}). Hypertension is a highly prevalent chronic medical condition and a strong modifiable risk factor for cardiovascular disease (CVD), as it attributes to more than $45\%$ of cardiovascular disease and $51\%$ of stroke deaths (\cite{WHO2013}). The risk of CVD in individuals rises sharply with increasing BP (\cite{EttehadEtAl2016, BundyEtAl2017, Collab2002, NavarEtAl2016, ClarkEtAl2019}). 

Continuous BP measurement has proven to be one of effective incident prevention. This implies that BP is the essential physiological indicator of human body. When the heart beats, it pumps blood to the arteries resulting in changes of BP during the process. When the heart contracts, BP in the vessels reaches its maximum, which is known as systolic BP (SBP). When the heart rests, BP reduces to its minimum, which is known as diastolic BP (DBP). 

Linear regression and polynomial regression analyses have been used in assessing the association between BP and risk factors contributing to various diseases (\cite{Koh2022}; \cite{Liu2022}; \cite{Yeo2022}). It is evident that the polynomial regression models fit the data accurately in some research studies due to its adaptability of nonlinearity property but face high order polynomial approximation. The fractional polynomials (FPs) proposed by \cite{RoystonAltman1994} act as a concise and accurate formulae for examining smooth relationships between response and predictors, and a compromise between precision and generalisability. FPs are parametric in nature and then intuitive for the interpretation of the analysis results. FP approach has clearly established a role in the nonlinear parametric methodology especially with application by clinicians from various research fields, such as obstetrics and gynecology (\cite{Tilling2014}), gene expression studies in clinical genetics (\cite{Tan2011}) and cognitive function of children (\cite{Ryoo2017}), and other medical applications (\cite{Wang2011}; \cite{Ravaghi2020}; \cite{Frangou2021} and among others). 

However, modelling conditional mean functions observes the partial view of a distribution of response variable, as the distributions of many response variables such as the BP measures are typically skew. Then `average’ BP may link to CVD but extremely high BP could explore CVD insight deeply and precisely. So, existing mean-based FP approaches for modelling the relationship between factors and BP cannot answer key questions in need. It is attractive to model conditional quantile functions with FPs that accommodates skewness very easily. Quantile regression, introduced by \cite{KoenkerBassett1978}, provides comprehensive relationship between the response variable and its predictors, such as median and extremely high BP measures may be often required in practical data analysis generally. 

\cite{Zhan2021} suggested quantile regression with FP as a suitable approach for an application, such as age-specific reference values of discrete scales, in terms of model consistency, computational cost and robustness. This approach is also used to derive reference curves and reference intervals in several applications (\cite{Chitty2003}; \cite{Bell2010}; \cite{Bedogni2012}; \cite{Kroon2017};  \cite{Casati2019}; \cite{Cai2020}; \cite{Loef2020}), which allow quantiles to be estimated as a function of covariates without requiring parametric distributional assumptions. This is essential for data that do not assume normality, linearity and constant variance. Recently, reasonable amount of nonlinear quantile regression analyses have been conducted in medical data analysis (\cite{Maidman2018}; \cite{HuangHan2023}; \cite{WuDupuis2023} and among others). 

However, Bayesian approach to quantile regression has advantages over the frequentist approach, as it can lead to exact inference in estimating the influence of risk factors on the upper quantiles of the conditional distribution of BP compared to the asymptotic inference of the frequentist approach (\cite{YuVanZhang2005}). It also provides estimation that incorporates parameter uncertainty fully (\cite{YuMoyeed2001}; \cite{YuVanZhang2005}). Some comparison studies have been conducted for both Bayesian and frequentist approaches, such as the analysis of risk factors for female CVD patients in Malaysia (\cite{Juhan2020}) and the analysis of risk factors of hypertension in South Africa (\cite{KuhudzaiEtAl2022}). The former revealed that the Bayesian approach has smaller standard errors than that of the frequentist approach. The latter also revealed that credible intervals of the Bayesian approach are narrower than confidence intervals of the frequentist approach. These findings suggest that the Bayesian approach provides more precise estimates than the frequentist approach. 

Variable selection in Bayesian quantile regression has been widely studied in the literature (\cite{LiEtAl2010}; \cite{Alhamzawi2012}; \cite{Alhamzawi2013}; \cite{ChenEtAl2013-BQR}; \cite{Adlouni2018}; \cite{Alhamzawi2019}; \cite{Mao2022} and among others). It plays an important role in building a multiple regression model, provides regularisation for good estimation of effects, and identifies important variables. \cite{BoveHeld2011} combine variable selection and 'parsimonious parametric modelling' of \cite{RoystonAltman1994} to formulate a Bayesian multivariate FP model with variable selection that efficiently selects best fitted FP model via stochastic search algorithm. However, In present, no research studies have been conducted for variable selection in Bayesian parametric nonlinear quantile regression for medical application even though there is a limited amount of studies in case of non-regularised models, such as mixed effect models (\cite{Wang2012}; \cite{YuYu2023}).

Therefore, in this paper, we explore a new quantile regression model using FPs and employ Bayesian variable selection with quantile-dependent prior for a more accurate representation of the risk factors on BP measures. The three-stage computational scheme of \cite{Mao2022} is employed as a variable selection method due to its fast convergence rate, low approximation error and guaranteed posterior consistency under model misspecification. So, we propose a Bayesian variable selection with nonlinear quantile regression model to assess how body mass index (BMI) among the United States (US) influences BP measures, including SBP and DBP. The objective of this paper is to examine a nonlinear relationship between BP measures and their risk factors across median and upper quantile levels. The dataset used in this paper is the 2007-2008 National Health and Nutrition Examination Survey (NHANES), including the information on BP measurements, body measures and sociodemographic questionnaires. 

The remainder of this paper is as follows. Section \ref{sec:methodology} presents the concept of FPs (\cite{RoystonAltman1994}), quantile regression (\cite{KoenkerBassett1978}) and Bayesian variable selection with quantile-dependent prior (\cite{Mao2022}). The details of the NHANES 2007-2008 dataset used for the analysis are provided in Section \ref{sec:Data}. Section \ref{sec:Results} applies the proposed method to the analysis of the NHANES 2007-2008 dataset, performs comparative analysis with two quantile regression methods and provides all the findings. Section \ref{sec:conclusion} concludes this paper.

\section{Methodology}\label{sec:methodology}
Regression analysis is a technique that quantifies the relationship between a response variable and predictors. Quantile regression, introduced by \cite{KoenkerBassett1978}, is a method to estimate the quantiles of a conditional distribution of a response variable and such it permits a more accurate portrayal of the relationship between the response variable and predictors. Unlike linear regression analysis, quantile regression analysis gives a better idea about distribution of the data because the latter is robust to outliers. 

\subsection{Quantile Regression}

Let $\tau$ be the proportion of a sample having data points below the quantile in $\tau$. Given a dataset, $\{x_i,y_i\}^n_{i=1}$ and fixed $\tau$, the $\tau^{th}$ quantile regression model is represented as 
\begin{align}
  y_i = x_i^T\beta(\tau) +\epsilon(\tau)_i\,, \quad i=1,\ldots,n\,, \label{eq:quantileeqf}
\end{align}
where $\tau$ is in the range between $0$ and $1$, and $\beta(\tau)$ is the vector of unknown parameters of interest and $\epsilon(\tau)$ is the model error term for the $\tau^{th}$ quantile. For the sake of notation simplification, we omit $\tau$ from these parameters. 

We wish to estimate the unknown parameters, $\beta $ as $\hat{ \beta}$ for each $\tau^{th}$ quantile, which can be done by minimising the check function over $\beta$:
\begin{equation}\label{minimisequant}
  \sum^{n}_{i=1} \rho_{\tau}(y_i-x^{T}_{i}\beta)\,,
\end{equation}
with the check function $\rho_{\tau}(\Delta)=\Delta \left[ \tau\cdot \I_{\Delta\geq 0}-(1-\tau)\cdot \I_{\Delta<0} \right]$ where $\I_{\Delta\geq 0}$ represents the value $1$ if $\Delta$ belongs to the set $[0,\infty)$, and the value $0$ otherwise. 

Minimising (\ref{minimisequant}) is same as maximising a likelihood function. An asymmetric Laplace distribution (ALD) is employed, which is the common choice for the quantile regression analysis (\cite{YuMoyeed2001,Yu2003}). We assume that $\epsilon_i\sim \mathcal{AL}(0,\sigma,\tau), i=1,\ldots,n$, where the $\mathcal{AL}$ is the ALD with its density 
  \begin{align*}
    f_{AL}(\epsilon_i) = \frac{\tau(1-\tau)}{\sigma}\exp\left\{-\frac{\rho_\tau(\epsilon_i) }{\sigma}\right\}\,.
  \end{align*}
Here, $\rho_\tau(\epsilon_i)$ denotes the usual check loss function of \cite{KoenkerBassett1978}.

We are interested in selecting a subset of important predictors which has adequate explanatory and predictive capability. One of the common procedures for simultaneously facilitating the parameter estimation and variable selection is to impose penalty function on the likelihood to arrive at the penalised loss function, 
\begin{equation}\label{minquant_penalised}
  \sum^{n}_{i=1} \rho_{\tau}(y_i-x^{T}_{i}\beta) + P(\beta,\delta)\,,
\end{equation}
which is minimised to obtain the $\tau^{th}$ quantile regression estimator. Here, $P(\beta,\delta)$ is a regularisation penalty function and $\delta$ is a penalty parameter that controls the level of sparsity. Typically, Bayesian regularised qauantile regression is formulated through the relationship between the check function and the ALD. 

Bayesian inference is one of the most popular approaches for the regression analysis since it provides with an entire posterior distribution of a parameter of interest as well as incorporation of parameter uncertainty and prior information about data. So, Bayesian analysis is preferable over frequentist analysis.

By using the identity of \cite{AndrewsEtAl1974}, 
\begin{align*}
  \exp(-|ab|) = \int^\infty_0 \frac{a}{\sqrt{2\pi v}} \exp\left\{ - \frac{1}{2} (a^2v + b^2v^{-1}) \right\} d\nu\,,
\end{align*}
for any $a,b>0$, letting $a=1/\sqrt{2\sigma}$ \& $b=\epsilon/\sqrt{2\sigma}$ and multiplying a factor of $\exp(-(2\tau-1)\epsilon/2\sigma)$, to express the probability density function (pdf) of the ALD errors as its scale mixture of Normals (SMN) representation,
\begin{align*}
  f_{AL}(\epsilon_i) = \int^\infty_0 \frac{1}{\sqrt{4\pi\sigma^3v_i}} \exp\left\{ - \frac{(\epsilon_i - (1-2\tau)v_i)^2}{4\sigma v_i} - \frac{\tau(1-\tau)v_i}{\sigma} \right\} dv_i\,,
\end{align*}
(\cite{KozumiKobayashi2011}). This representation can be utilised to enable facilitation of Gibbs sampling algorithms (\cite{KozubowskiEtAl2001}; \cite{GeraciBottai2007}; \cite{KozumiKobayashi2011}; \cite{ChenEtAl2013-BQR}). 

Rather than the standard linear model, we will be using the FP model to develop the nonlinear model under Bayesian quantile regression and variable selection. 

\subsection{Fractional Polynomials}

\cite{BoxTidwell1962} introduced the transformation now known as the Box-Tidwell transformation,
   \begin{equation}\label{EQN1}
    x^{(a)} = 
    \begin{cases}
    x^a, & \text{if} \quad a\neq 0\,, \nonumber \\
    \log(x), & \text{if} \quad a = 0\,, 
    \end{cases}
  \end{equation}
where $a$ is a real number.
\cite{RoystonAltman1997} extend the classical polynomials to a class which they called FPs.

An FP of degree $m$ with powers $p_1\leq\ldots\leq p_m$ and respective coefficients $\alpha_1,\ldots,\alpha_m$ is 
  \begin{align*}
    f^m(x;\bm{\alpha},\bm{p}) = \sum^m_{j=1}\alpha_j h_j(x)\,,
    \end{align*}
where $h_0(x)=1$ and
\begin{numcases}{h_j(x) =} 
    \text{$x^{(p_j)}$}, & \text{if $p_j\neq p_{j-1}$}\,,\nonumber\\
    \text{$h_{j-1}(x)$}\log(x), & \text{if $p_j = p_{j-1}$}\,, \label{EQN2}
\end{numcases}
where $j=1\ldots,m$.
Note that the definition $h_j(x)$ allows the repeated powers.
The bracket around the exponent denote the Box-Tidwell transformation (\ref{EQN1}).
For $m\leq 3$, \cite{RoystonAltman1994} constrained the set of possible powers $p_j$ to the set 
  \begin{align}
    \mathcal{S}=\left\{-2,-1,-\frac{1}{2},0,\frac{1}{2},1,2,3\right\}\,, \label{EQN4}
  \end{align}
which encompasses the classical polynomial powers $1,2,3$ but also offers square roots and reciprocals. \cite{RoystonSauerbrei2008} argue that this set is sufficient to approximate all powers in internals $[-2,3]$. The simple example of the FP model is as follows. An FP with $m=3$ powers and its power vector $\bm{p}= (p_1,p_2,p_3)=\left(-\frac{1}{2},2,2\right)$ would be 
  \begin{align*}
    f^3(x;\bm{\alpha},\bm{p}) = \alpha_1 x^{-1/2} +\alpha_2 x^2 + \alpha_3 x^2\log(x)\,,
  \end{align*}
where the last term reflects the repeated power $2$.
  
Generalisation to the case of multiple predictors:
  \begin{align}
    \eta(\bm{x}) = \sum^k_{l=1} f_l^{m_l}(x_l;\alpha_l,p_l) = \sum^k_{l=1}\sum^{m_l}_{j=1} \alpha_{lj}h_{lj}(x_l)\,. \label{EQN3}
  \end{align}
This is called the multiple FP model. Suppose we continue examining $k$ continuous predictors $x_1,\ldots,x_k$ and content themselves with a maximum degrees of $m_{max}\leq 3$ for each $f_l^{m_l}$, for instance, $0\leq m_l\leq m_{max}$ for $l=1,\ldots,k$, where $m_l=0$ denotes the omission of $x_l$ from the model. From the powers set $\mathcal{S}$, $m_l$ powers are chosen, which need not be different due to the inclusion of logarithmic terms for repeated powers (\ref{EQN2}), we now employ the $\tau^{th}$ nonlinear quantile regression with the SMN representation of the ALD errors, 
\begin{align}\label{eq:QR-FP}
    \bm{y} = \bm{B}\bm{\beta}+\theta_1 \bm{v} + \sqrt{\theta_2 \bm{v} \sigma^2} \bm{z}\,,
  \end{align}

where the $(n\times D)$-matrix $\bm{B}$ is a function of predictors $x_l$ of the $i^{th}$ observations ($i=1,\ldots,n$), $\bm{v}=(v_1,\ldots,v_n)^T$ is a vector of exponential random variables with a rate of $\frac{\tau(1-\tau)}{\sigma}$, $\bm{z}=(z_1,\ldots,z_n)^T$ is a vector of standard Normal random variables and $z_i \ci v_i$ for $i=1,\ldots,n$, $\theta_1 = \frac{1-2\tau}{\tau (1-\tau)}$ and $\theta_2 = \frac{2}{\tau(1-\tau)}$. Each entry of matrix $\bm{B}$ is a vector, $\bm{B}_{id} = \bm{B}(x_{id}) = (\alpha_{l1}h_{l1}(x_{il}),\ldots, \alpha_{lm_l}h_{lm_l}(x_{il}))^T$, for $i=1,\ldots,n$, $l=1,\ldots,k$ and $d=1,\ldots,D$. 

A special way of defining the matrix $\bm{B}$ is through the use of FPs. In this case, the basis function $B(x_l)$ is chosen as the transformation $h_{lj}$ in (\ref{EQN3}) ($j=1,\ldots,m_l$) and the unknown parameter $\bm{\beta}=(\bm{\alpha}_1,\ldots,\bm{\alpha}_{k})^T$, where $\bm{\alpha}_l=(\alpha_{l1},\ldots, \alpha_{lm_l})$ for $l=1,\ldots,k$. The transformation $h_j$ are determined by the power vector $\bm{p}_1,\ldots, \bm{p}_k$ through their definition (\ref{EQN2}). Note that the $\bm{p}_l$ is empty if the predictor $x_l$ is not included in the model ($m_l=0$).

\subsection{Bayesian Approach and Variable Selection}

Given the model in (\ref{eq:QR-FP}), the likelihood function conditional on $\bm{\beta}, \sigma, \bm{v}=(v_1,\ldots,v_n)^T$ can be written as 
\begin{align*}
  f(\bm{y}|\bm{\beta}, \sigma, \bm{v},\bm{B}) = \prod^n_{i=1} \frac{1}{\sqrt{4\pi\sigma^3 v_i}} \exp\left\{ - \frac{(y_i - \bm{B}(x_i)^T\bm{\beta} - (1-2\tau) v_i)^2}{4\sigma v_i} -\frac{\tau(1-\tau)v_i}{\sigma} \right\}\,.
\end{align*}

We employ the three-stage algorithm of \cite{Mao2022} for Bayesian nonlinear quantile regression with variable selection. It can be summarised, as follows.

The first stage is the expectation-maximisation algorithm consisting of two main steps: the E-step and the M-step. \cite{Dempster1977} proposed the EM algorithm, which is a statistical simulation method and it aims to solve the complex data analysis problem with missing data. 

Suppose the complete data $(\bm{y},\bm{v})$ is composed of the observed data $\bm{y}=(y_1,\ldots,y_n)^T$ and missing data $\bm{v}=(v_1,\ldots,v_n)^T$, whereas $\bm{B}(x_i)$, $i=1,\ldots,n,$ are treated as a function of fixed predictors. Maximum likelihood estimates (MLE) can be obtained by maximising log-likelihood function $\log f( \bm{\beta},\sigma|\bm{y},\bm{v})$ of the complete data. EM algorithm has the following two steps: Expectation step (E step) and Maximum step (M step).
\begin{itemize}[label={}]
  \item{[E step]} Given initial values of $\bm{\beta}^{(0)}$ and $\sigma^{(0)}$, we denote $\bm{\beta}^{(q-1)}$ and $\sigma^{(q-1)}$ as the $(q-1)^{th}$ iteration value of parameters $\bm{\beta}$ and $\sigma$ in the EM algorithm, and we define the mathematical expectation of the complete data as a Q-function
  \begin{align*}
    Q(\bm{\beta},\sigma|\bm{y},\bm{\beta}^{(q-1)}, \sigma^{(q-1)})=\mathbb{E}_{\bm{y},\bm{\beta}^{(q-1)}, \sigma^{(q-1)}}[ \log f( \bm{\beta},\sigma|\bm{y},\bm{v})]\,.
  \end{align*}
  \item{[M step]} We obtain the updated values of $\bm{\beta}^{(q)}$ and $\sigma^{(q)}$ by maximising $Q(\bm{\beta},\sigma|\bm{y},\bm{\beta}^{(q-1)}, \sigma^{(q-1)})$ over parameters $\bm{\beta}$ and $\sigma$:
  \begin{align*}
  \bm{\beta}^{(q)} &= (\bm{B}^T\bm{W}^{(q-1)}\bm{B})^{-1} \bm{B}^T\bm{W}^{(q-1)}(\bm{y}-\theta_1\bm{\Delta 3})\,,
    \intertext{where $\bm{\Delta 3} = \left(\left|y_1-\bm{B}(x_1)^T\bm{\beta}^{(q-1)}\right|, \ldots, \left|y_n-\bm{B}(x_n)^T\bm{\beta}\right|^{(q-1)}\right)^T$ and $\bm{W}^{(q-1)}=\text{diag}(1/\Delta 3_1,\ldots,1/\Delta 3_n)$, and }
    \sigma^{(q)} &=\frac{1}{2(3n+2)} \left\{ \sum^n_{i=1} \Delta 2_i + \sum^n_{i=1} \frac{(y_i-\bm{B}(x_i)^T\bm{\beta}^{(q)})^2}{\Delta 3_i} -2\theta_1\sum^n_{i=1} (y_i-\bm{B}(x_i)^T\bm{\beta}^{(q)}) \right\}\,,
    \intertext{where $\Delta 2_i =\left|y_i-\bm{B}(x_i)^T\bm{\beta}^{(q-1)}\right|+2\sigma^{(q-1)}$ for $i=1,\ldots,n$.}
  \end{align*}
\end{itemize}
Repeat E-step and M-step until it meets the required condition, then the final iteration values of the EM algorithm are set as the posterior modes of $\bm{\beta}$ and $\sigma$, denoted by $\Tilde{\bm{\beta}}$ and $\Tilde{\sigma}$, respectively. 

The second stage is the Gibbs sampling algorithm. The quantile-specific Zellner's $g$-prior (\cite{AlhamzawiYu2013}) is used for the prior specification and it is given by 
\begin{align}\label{eq:prior}
  \bm{\beta}| \sigma,\bm{V},\bm{B} \sim N\left(0, 2\sigma g \bm{\Sigma}_v^{-1} \right) \quad \text{and} \quad p(\sigma) \propto \frac{1}{\sigma}\,,
\end{align}
where $N(\cdot)$ is the multivariable Normal distribution, $g$ is a scaling factor, $\bm{V} = \text{diag}(1/v_1,\ldots,1/v_n)$ and $\bm{\Sigma}_v = \bm{B}^T\bm{V}\bm{B}$. This prior specification has an advantage, as it contains information that is dependent upon the quantile levels, which increases posterior inference accuracy. 

Given the posterior modes, $\Tilde{\bm{\beta}}$ and $\Tilde{\sigma}$ as the starting value, we denote $\bm{\beta}^{(r-1)}$ and $\sigma^{(r-1)}$ as the $(r-1)th$ iteration value of parameters $\bm{\beta}$ and $\sigma$ in the Gibbs sampling algorithm. 

\begin{itemize}

  \item Sample $v_i^{(r)}$ from 
  $$p(v_i) \sim GIG\left(0, \frac{1}{2\sigma}, \frac{(y_i-\bm{B}(x_i)^T\bm{\beta})^2 + \frac{1}{g} \bm{\beta}^T\bm{B}(x_i)\bm{B}(x_i)^T\bm{\beta} }{2\sigma} \right),$$
  based on $\bm{\beta}^{(r-1)}$ and $\sigma^{(r-1)}$ for $i=1,\ldots,n$ and $GIG(0,c,d)$ is the generalised inverse Gaussian with its density
  \begin{align*}
    f_{\text{GIG}}(v) = \frac{1}{2K_0(\sqrt{cd})} v^{-1} \exp\left(-\frac{1}{2} (cv+dv^{-1}) \right), \quad v>0\,,
  \end{align*}
where $K(\cdot)$ is the modified Bessel function of the third kind (\cite{BarndorffNielsen2001}).

  \item Sample $\sigma^{(r)}$ from 
  $$p(\sigma|\bm{y},\bm{v}^{(r)}) \sim IG\left(\frac{3n}{2}, \frac{1}{4} (\bm{y}-\theta_1\bm{v})^T \bm{V}\bm{H}_v (\bm{y}-\theta_1\bm{v}) + \frac{2}{\theta_2} \sum^n_{i=1}v_i \right),$$
  where $IG(\cdot)$ is the inverse Gamma distribution, $\bm{H}_v=\bm{I}_n - \frac{g}{g+1}\bm{B}\bm{\Sigma}_v^{-1}\bm{B}^T\bm{V}$.
  
  \item Sample $\bm{\beta}^{(r)}$ from 
  $$p(\bm{\beta}|\bm{y},\bm{v}^{(r)},\sigma^{(r)}) \sim N\left(\frac{g}{g+1}\bm{\Sigma}^{-1}_v\bm{B}^T\bm{V}(\bm{y}-\theta_1\bm{v}), \frac{2\sigma g}{g+1} \bm{\Sigma}^{-1}_v \right)\,.$$
  
  \item Calculate the important weights 
  $$w^{(r)} = \frac{p(\bm{\beta}^{(r)},\sigma^{(r)},\bm{v}^{(r)}|\bm{y})}{p(\bm{\beta}^{(r)}|\sigma^{(r)},\bm{b}^{(r)},\bm{y})p(\sigma^{(r)}|\bm{v}^{(r)},\bm{y})p(\bm{v}^{(r)})}\,,$$
  based on $\bm{v}^{(r)}, \sigma^{(r)}, \bm{\beta}^{(r)}$. This is to adjust for the GIG approximation of the marginal posterior of $\bm{v}$ given $\bm{y}$, which is given by its unnormalised density function
  \begin{align*}
    \pi(\bm{v}|\bm{y}) \propto \frac{p(\bm{v}|\Tilde{\bm{\beta}},\Tilde{\sigma},\bm{y})}{p(\Tilde{\bm{\beta}}|\bm{y},\bm{v}.\Tilde{\sigma})p(\Tilde{\sigma}|\bm{y},\bm{v})}\,,
  \end{align*}
  where $p(\bm{v}|\Tilde{\bm{\beta}},\Tilde{\sigma},\bm{y})$ is an importance sampling desnity in the importance smapling algorithm. 
  The importance weights will be used to determine the acceptance probability of each $\{\bm{\beta}^{(r)}, \sigma^{(r)}, \bm{v}^{(r)}\}$.
\end{itemize}

The algorithm iterates until it reaches the final MCMC iteration indexed at R and discard the burn-in period. 

Finally, the third stage is the important re-weighting step. The $S$ samples are drawn from the importance weights without replacement where $S<R$ is the number of importance weighting steps. A random indicator vector $\bm{\gamma}=(\gamma_1,\ldots,\gamma_D)^T$ is introduced to the nonlinear model
\begin{align*}
  \bm{M}_{\bm{\gamma}}: \bm{y} = \bm{B}_{\bm{\gamma}}\bm{\beta} + \bm{\epsilon}\,,
\end{align*}
where $\bm{B}_{\bm{\gamma}}$ is the $(n\times D_{\bm{\gamma}})$ matrix consisting of important predictors and $\bm{\beta}_{\bm{\gamma}}$ of length $D_{\bm{\gamma}}$ is the non-zero parameter vector. The same prior specification in (\ref{eq:prior}) is employed along with a prior on $\gamma_d$, $d=1,\ldots,D$, and a beta prior on $\pi$:
\begin{align*}
  p(\bm{\gamma}|\pi)\propto \pi^{\sum^D_{d=1}\gamma_d} (1-\pi)^{D-\sum^D_{d=1}\gamma_d} \quad \text{and} \quad p(\pi)\sim \text{Beta}\left(\frac{1}{2}, \frac{1}{2} \right)\,,
\end{align*}
where $\pi\in[0,1]$ is the prior probability of randomly including predictor in the model. Then $\pi$ is marginalised out from $p(\bm{\gamma}|\pi)$ resulting as
\begin{align*}
  p(\bm{\gamma})\propto \text{Beta} \left(\sum^D_{d=1}\gamma_d +\frac{1}{2}, D - \sum^D_{d=1}\gamma_d +\frac{1}{2} \right)\,.
\end{align*}
The marginal likelihood of $\bm{y}$ under the model $\bm{M}_{\bm{\gamma}}$ is then obtained by integrating out $\bm{\beta}$ and $\sigma$ resulting as
\begin{align*}
  p(\bm{y}|\bm{\gamma},\bm{v}) \sim t_{2n} \left((1-2\tau)\bm{v}, \frac{4\sum^n_{i=1}v_i}{\sigma\theta_2} \left(\bm{V} - \frac{g}{g+1}\bm{V}\bm{B}_{\bm{\gamma}}\bm{\Sigma}_v (\bm{\gamma})^{-1} \bm{B}_{\bm{\gamma}}^T\bm{V} \right)^{-1} \right)\,,
\end{align*}
where $t_{2n}(\cdot)$ is the multivariate Student t-distribution with $2n$ degrees of freedom. The posterior probability of $\bm{M}_{\bm{\gamma}}$ is therefore given by $p(\bm{\gamma}|\bm{y},\bm{v})\propto p(\bm{y}|\bm{\gamma},\bm{v}) p(\bm{\gamma})$. Lastly, the independent samples of $\bm{v}$ from the second stage algorithm are drawn based on the $S$ samples and the important re-weighting step is iterated until the $S$ samples of $\bm{\gamma}$ are obtained. Then the posterior inclusion probability is estimated, as follows
\begin{align*}
  \hat{p}(\gamma_d=1|\bm{y},\bm{v}) = \frac{1}{\Tilde{S}} \sum^{\Tilde{S}}_{s=1} \gamma_d^{(s)}\,, \quad d=1,\ldots, D\,,
\end{align*}
where $\Tilde{S}$ is the number of iterations after discarding the burn-in period.

\section{Data Preparation and Data Analysis}\label{sec:Data}

This study is based on the data of the National Health and Nutrition Examination Survey (NHANES) during 2007-2008. The survey conducted by the National Center for Health Statistics of the Centers for Disease Control and Prevention used a complex, stratified, multistage sampling design to select a representative sample of noninstitutionalized population in the United Status civilians to participate in a series of comprehensive health-related interviews and examinations. In total, 12,943 people participated in NHANES 2007-2008. 

The study variables included SBP and DBP as the response variables. The BP measurements were taken as follows. After a resting period of 5 minutes in a sitting position and determination of maximal inflation level, three consecutive BP readings were recorded. A fourth reading was recorded if a BP measurement is interrupted or incomplete. All the results were taken in Mobile Examination Center. The BP measurements are essential for hypertension screening and disease management, since hypertension is an important risk factor for cardiovascular and renal disease. Then in this study, SBP and DBP were selected as response variables where each was averaged over the second and third readings. Predictor variables were BMI, age, ethnicity, gender and marital status. 

We initially included 9,762 participants who have completed both BP and body measure examinations in the study. From 9,762 participants, we exclude those who had not underwent examinations. Then among the remaining 4,612 participants, we further excluded those who refused to reveal their marital status. Finally, 4,609 participants were included for analysis in this study. 

The NHANES protocols were approved by the National Center for Health Statistics research ethics review boards, and informed consent was obtained from all participants. The research adhered to the tenets of the Declaration of Helsinki.

The R version $4.2.2$ was used to conduct both statistical and Bayesian analyses. Both 'quantreg' and 'Brq' R packages was employed to fit the frequentist and Bayesian approaches of the quantile regression model with FPs, respectively. The source R code was provided from the main author to fit the Bayesian quantile regression with variable selection and FPs via the three-stage algorithm.

This study considers two quantile models at the $50^{th}$, $75^{th}$ and $95^{th}$ percentiles. When modelling hypertension, it is preferable to model both median and extremely high values of SBP and DBP, which corresponds to the median and upper distributions of SBP and DBP, respectively (\cite{KuhudzaiEtAl2022}). The following two quantile models will be used for the analysis for the fixed $\tau$ value:
\begin{align*}
  &\text{SBP}_i = \text{BMI}_i\beta_1 + \text{BMI}^{0.5} \beta_2 + \text{Age}_i\beta_3 + \text{Age}_i^{0.5}\beta_4 + \text{Ethnicity}_i\beta_5 + \text{Gender}_i\beta_6 + \text{MaritalStatus}_i\beta_7\,, \\
  &\text{DBP}_i = \text{BMI}_i\beta_1 + \text{BMI}^{0.5} \beta_2 + \text{Age}_i\beta_3 + \text{Age}_i^{0.5}\beta_4 + \text{Ethnicity}_i\beta_5 + \text{Gender}_i\beta_6 + \text{MaritalStatus}_i\beta_7\,,
\end{align*}
for $i=1,\ldots, 4609$.

The power of $0.5$ was chosen for continuous variables, including BMI and age. The remaining variables were linear because they are categorical. Similar fractional polynomial models were employed to model BP within the linear regression framework (\cite{Dong2016}, \cite{Takagi2013}, \cite{Thompson2009} and among others). 

\section{Results}\label{sec:Results}

\begin{table}[htbp]
\caption{SBP among United Status Adults by BMI and Sociodemographic Characteristics.}
\label{tab:SBPdes}
\tabcolsep=0pt
\begin{tabular*}{\textwidth}{@{\extracolsep{\fill}}llrrr@{\extracolsep{\fill}}}
\toprule
\multicolumn{2}{l}{} & Normal BP & Pre- & Hypertension \\
\multicolumn{2}{l}{} & (< 120 mmHg) & Hypertension & ($\geq$ 140 mmHg) \\
\multicolumn{2}{l}{} & & (120-139 mmHg) & \\
\cline{3-5}
BMI & Underweight   & 37 (56.92\%) &16 (24.62\%) & 12 (18.46\%) \\
  & Healthy     & 734 (60.31\%) &343 (28.18\%) & 140 (11.50\%) \\
  & Overweight    & 781 (49.49\%) & 565 (35.80\%) & 232 (14.70\%) \\
  & Obese      & 415 (41.71\%) & 414 (41.61\%) & 166 (16.68\%) \\
  & Very obese    & 201 (42.68\%) & 187 (39.70\%) & 83 (17.62\%) \\
  & Morbidly obese  & 106 (37.46\%) & 116 (40.99\%) & 61 (21.55\%) \\
\multicolumn{2}{l}{P-value (Cramer's V value} & \multicolumn{3}{c}{P-value < 0.01 (0.1106)} \\
\\
Age & 20-29 years & 493 (73.36\%) & 164 (24.40\%) & 15 (2.23\%) \\
  & 30-39 years & 543 (65.66\%) & 251 (30.35\%) & 33 (3.99\%) \\
  & 40-49 years & 460 (55.89\%) & 285 (34.63\%) & 78 (9.48\%) \\
  & >50 years  & 778 (34.02\%) & 941 (41.15\%) & 568 (24.84\%) \\
\multicolumn{2}{l}{} & \multicolumn{3}{c}{P-value < 0.01 (0.2535)} \\  
\\
Ethnicity & Mexican American    & 456 (54.29\%) & 279 (33.21\%) & 105 (12.50\%) \\
     & Other Hispanic     & 286 (53.16\%) & 186 (34.57\%) & 66 (12.27\%) \\
     & Non-Hispanic white   & 1006 (47.61\%) & 793 (37.53\%) & 314 (14.86\%) \\
     & Non-Hispanic black   & 425 (45.31\%) & 324 (34.54\%) & 189 (20.15\%) \\
     & Other non-Hispanic race & 101 (56.11\%) & 59 (32.78\%) & 20 (11.11\%) \\
\multicolumn{2}{l}{} & \multicolumn{3}{c}{P-value < 0.01 (0.0665)} \\  
\\
Gender & Male & 999 (43.28\%) & 957 (41.46\%) & 352 (15.25\%) \\
   & Female & 1275 (55.41\%) & 684 (29.73\%) & 342 (14.86\%) \\
\multicolumn{2}{l}{} & \multicolumn{3}{c}{P-value < 0.01 (0.1310)} \\
\\
Marital & Married       & 1219 (48.39\%) & 927 (36.80\%) & 373 (14.81\%) \\
 Status & Widowed       & 84 (30.11\%) & 103 (36.92\%) & 92 (32.97\%) \\
     & Divorced      & 226 (44.14\%) & 182 (35.55\%) & 104 (20.31\%) \\
     & Separated      & 89 (52.05\%) & 57 (33.33\%) & 25 (14.62\%) \\
     & Never married    & 468 (58.87\%) & 256 (32.20\%) & 71 (8.93\%) \\
     & Living with partner & 188 (56.46\%) & 116 (34.83\%) & 29 (8.71\%) \\
\multicolumn{2}{l}{} & \multicolumn{3}{c}{P-value < 0.01 (0.1251)} \\  
\bottomrule
\end{tabular*}
\end{table}

\begin{table}[htbp]
\caption{DBP among United Status Adults by BMI and Sociodemographic Characteristics.}
\label{tab:DBPdes}
\tabcolsep=0pt
\begin{tabular*}{\textwidth}{@{\extracolsep{\fill}}llrrr@{\extracolsep{\fill}}}
\toprule
\multicolumn{2}{l}{} & Normal BP & Pre- & Hypertension \\
\multicolumn{2}{l}{} & (< 80 mmHg) & Hypertension & ($\geq$ 90 mmHg) \\
\multicolumn{2}{l}{} & & (80-89 mmHg) & \\
\cline{3-5}
BMI & Underweight   & 49 (75.38\%) & 12 (18.46\%) & 4 (6.15\%) \\
  & Healthy     & 1025 (84.22\%) & 148 (12.16\%) & 44 (3.62\%)\\
  & Overweight    & 1265 (80.16\%) & 243 (15.40\%) & 70 (4.44\%)\\
  & Obese      & 772 (77.59\%) & 168 (16.88\%) & 55 (5.53\%)\\
  & Very obese    & 356 (75.58\%) & 78 (16.56\%) & 37 (7.86\%)\\
  & Morbidly obese  & 217 (76.68\%) & 47 (16.61\%) & 19 (6.71\%) \\
\multicolumn{2}{l}{P-value (Cramer's V value} & \multicolumn{3}{c}{P-value < 0.01 (0.0587)}  \\
\\
Age & 20-29 years & 619 (92.11\%) & 47 (6.99\%) & 6 (0.89\%) \\
  & 30-39 years & 681 (82.35\%) & 118 (14.27\%) & 28 (3.39\%) \\
  & 40-49 years & 584 (70.96\%) & 173 (21.02\%) & 66 (8.02\%) \\
  & >50 years  & 1800 (78.71\%) & 358 (15.65\%) & 129 (5.64\%) \\
\multicolumn{2}{l}{} & \multicolumn{3}{c}{P-value < 0.01 (0.1118)} \\  
\\
Ethnicity & Mexican American    & 699 (83.21\%) & 116 (13.81\%) & 25 (2.98\%) \\
     & Other Hispanic     & 444 (82.53\%) & 70 (13.01\%) & 24 (4.46\%) \\
     & Non-Hispanic white   & 1687 (79.84\%) & 327 (15.48\%) & 99 (4.69\%) \\
     & Non-Hispanic black   & 711 (75.80\%) & 154 (16.42\%) & 73 (7.78\%) \\
     & Other non-Hispanic race & 143 (79.44\%) & 29 (16.11\%) & 8 (4.44\%) \\
\multicolumn{2}{l}{} & \multicolumn{3}{c}{P-value < 0.01 (0.0569)} \\  
\\
Gender & Male & 1732 (75.04\%) & 423 (18.33\%) & 153 (6.63\%) \\
   & Female & 1952 (84.83\%) & 273 (11.86\%) & 76 (3.30\%) \\
\multicolumn{2}{l}{} & \multicolumn{3}{c}{P-value < 0.01 (0.1244)} \\
\\
Marital & Married       & 2017 (80.07\%) & 385 (15.28\%) & 117 (4.64\%) \\
 Status & Widowed       & 231 (82.80\%) & 38 (13.62\%) & 10 (3.58\%)\\
     & Divorced      & 386 (75.39\%) & 87 (16.99\%) & 39 (7.62\%)\\
     & Separated      & 133 (77.78\%) & 26 (15.20\%) & 12 (7.02\%) \\
     & Never married    & 656 (82.52\%) & 103 (12.96\%) & 36 (4.53\%)\\
     & Living with partner & 261 (78.38\%) & 57 (17.12\%) & 15 (4.50\%)\\
\multicolumn{2}{l}{} & \multicolumn{3}{c}{P-value = 0.0516 (0.0444)} \\  
\bottomrule
\end{tabular*}
\end{table}

\begin{table}[htbp]
\caption{One Frequentist and Two Bayesian Quantile Regression Analyses for Relationship between SBP and Risk Factors.}
\label{tab:SBPest}
\tabcolsep=0pt
\begin{tabular*}{\textwidth}{@{\extracolsep{\fill}}lcrrr@{\extracolsep{\fill}}}
\toprule
\multicolumn{2}{l}{Quantile Regression} & & & \\
\cline{1-2}
      & \multicolumn{1}{c}{$\tau$} & \multicolumn{1}{c}{0.50} & \multicolumn{1}{c}{0.75} & \multicolumn{1}{c}{0.95} \\
BMI      & & -2.856 (-3.278, -2.280) & -2.198 (-3.040, -1.715) & -2.024 (-3.141, -0.798) \\
BMI${}^{0.5}$ & & 36.085 (29.932, 40.529) & 29.210 (23.907, 38.130) & 29.113 (15.239, 42.302) \\
Age      & & 0.510 (0.130, 0.785) & 0.317 (-0.003, 0.885) & 0.710 (-0.220, 1.630)\\
Age${}^{0.5}$ & & -1.758 (-5.430, 3.339) & 3.297 (-4.116, 7.654) & 2.300 (-9.906, 14.672) \\
Ethnicity   & & 0.626 (0.154, 1.040) & 0.995 (0.366, 1.495) & 1.214 (0.199, 2.642) \\
Gender     & & -4.323 (-5.302, -3.512) & -3.813 (-5.231, -2.506) & -3.278 (-6.147, -0.762) \\
Marital Status & & 0.894 (0.612, 1.155) & 1.327 (0.916, 1.746) & 1.400 (0.650, 2.037) \\
\hline

\multicolumn{2}{l}{Bayesian Quantile} & & & \\
\multicolumn{2}{l}{Regression} & & & \\
\cline{1-2}
      & \multicolumn{1}{c}{$\tau$} & \multicolumn{1}{c}{0.50} & \multicolumn{1}{c}{0.75} & \multicolumn{1}{c}{0.95} \\

BMI      & & -2.818 (-3.208, -2.447) & -2.255 (-2.669, -1.889) & -2.120 (-2.603, -1.685) \\
BMI${}^{0.5}$ & & 35.628 (31.653, 39.794) & 29.825 (25.763, 34.419) & 30.191 (25.146, 35.809) \\
Age      & & 0.484 (0.233, 0.734) & 0.364 (0.103, 0.664) & 0.768 (0.428, 1.142) \\
Age${}^{0.5}$ & & -1.366 (-4.737, 2.002) & 2.735 (-1.237, 6.249) & 1.446 (-3.550, 6.077) \\
Ethnicity   & & 0.640 (0.288, 0.979) & 0.957 (0.561, 1.359) & 1.341 (0.839, 1.829) \\
Gender     & & -4.376 (-5.138, -3.645) & -3.809 (-4.784, -2.823) & -3.346 (-4.397, -2.190) \\
Marital Status & & 0.888 (0.656, 1.125) & 1.347 (1.055, 1.637) & 1.354 (1.041, 1.649) \\

\hline

\multicolumn{2}{l}{Bayesian Quantile } & & & \\
\multicolumn{2}{l}{Regression Fractional } & & & \\
\multicolumn{2}{l}{Polynomials \&} & & & \\
\multicolumn{2}{l}{Variable Selection} & & & \\
\cline{1-2}
      & \multicolumn{1}{c}{$\tau$} & \multicolumn{1}{c}{0.50} & \multicolumn{1}{c}{0.75} & \multicolumn{1}{c}{0.95} \\
BMI      & & -2.812 (-3.164, -2.468) & -2.581 (-2.974, -2.168) & -2.426 (-2.813, -2.027) \\
BMI${}^{0.5}$ & & 35.547 (31.789, 39.269) & 33.335 (28.817, 37.747) & 33.335 (28.815, 37.784) \\
Age      & & 0.459 (0.226, 0.680) & 0.537 (0.274, 0.806) & 0.945 (0.643, 1.256) \\
Age${}^{0.5}$ & & -1.129 (-4.197, 2.029) & -0.051 (-3.717, 3.536) & -1.382 (-5.473, 2.680) \\
Ethnicity   & & 0.571 (0.258, 0.898) & 0.843 (0.484, 1.212) & 1.152 (0.753, 1.616) \\
Gender     & & -4.577 (-5.300, -3.899) & -4.291 (-5.053, -3.518) & -4.343 (-5.301, -3.351) \\
Marital Status & & 0.828 (0.632, 1.033) & 1.139 (0.893, 1.381) & 1.331 (1.052, 1.617) \\
 
\bottomrule
\end{tabular*}
\end{table}

\begin{table}[htbp]
\caption{One Frequentist and Two Bayesian Quantile Regression Analyses for Relationship between DBP and Risk Factors.}
\label{tab:DBPest}
\tabcolsep=0pt
\begin{tabular*}{\textwidth}{@{\extracolsep{\fill}}lcrrr@{\extracolsep{\fill}}}
\toprule
\multicolumn{2}{l}{Quantile Regression} & & & \\
\cline{1-2}
      & \multicolumn{1}{c}{$\tau$} & \multicolumn{1}{c}{0.50} & \multicolumn{1}{c}{0.75} & \multicolumn{1}{c}{0.95} \\
BMI      & & 1.174 (0.705, 1.496) & 0.761 (0.507, 1.096) & 0.582 (0.022, 1.572) \\
BMI${}^{0.5}$ & & -12.200 (-15.675, -7.071) & -7.179 (-10.821, -4.242) & -3.995 (-13.869, 2.247) \\
Age      & & -2.266 (-2.477, -1.979) & -2.018 (-2.252, -1.832) & -1.852 (-2.418, -1.418) \\
Age${}^{0.5}$ & & 31.329 (27.308, 34.170) & 28.298 (25.758, 31.451) & 26.918 (21.199, 34.557) \\
Ethnicity   & & 0.561 (0.203, 0.841) & 0.712 (0.411, 1.030) & 1.264 (0.345, 2.013) \\
Gender     & & -3.345 (-4.160, -2.651) & -3.619 (-4.337, -2.976) & -4.592 (-5.769, -3.047) \\
Marital Status & & 0.210 (-0.041, 0.448) & 0.368 (0.171, 0.549) & 0.466 (0.143, 0.934) \\
\hline

\multicolumn{2}{l}{Bayesian Quantile} & & & \\
\multicolumn{2}{l}{Regression} & & & \\
\cline{1-2}
      & \multicolumn{1}{c}{$\tau$} & \multicolumn{1}{c}{0.50} & \multicolumn{1}{c}{0.75} & \multicolumn{1}{c}{0.95} \\

BMI      & & 1.153 (0.836, 1.433) & 0.798 (0.539, 1.056) & 0.656 (0.345, 0.974) \\
BMI${}^{0.5}$ & & -11.923 (-15.007, -8.505) & -7.554 (-10.406, -4.748) & -4.624 (-7.981, -1.332) \\
Age      & & -2.253 (-2.431, -2.058) & -2.040 (-2.224, -1.863) & -1.870 (-2.064, -1.663) \\
Age${}^{0.5}$ & & 31.131 (28.434, 33.566) & 28.594 (26.243, 31.077) & 27.176 (24.467, 29.773) \\
Ethnicity   & & 0.536 (0.291, 0.777) & 0.706 (0.455, 0.966) & 1.328 (0.981, 1.667) \\
Gender     & & -3.391 (-3.999, -2.778) & -3.635 (-4.169, -3.109) & -4.498 (-5.086, -3.924) \\
Marital Status & & 0.220 (0.030, 0.408) & 0.374 (0.222, 0.533) & 0.484 (0.304, 0.667) \\

\hline

\multicolumn{2}{l}{Bayesian Quantile } & & & \\
\multicolumn{2}{l}{Regression Fractional } & & & \\
\multicolumn{2}{l}{Polynomials \&} & & & \\
\multicolumn{2}{l}{Variable Selection} & & & \\
\cline{1-2}
      & \multicolumn{1}{c}{$\tau$} & \multicolumn{1}{c}{0.50} & \multicolumn{1}{c}{0.75} & \multicolumn{1}{c}{0.95} \\
BMI      & & 1.101 (0.823, 1.381) & 0.808 (0.568, 1.041) & 0.874 (0.584, 1.147) \\
BMI${}^{0.5}$ & & -11.299 (-14.374, -8.289) & -7.620 (-10.207, -4.940) & -7.217 (-10.158, -4.080) \\
Age      & & -2.217 (-2.397, -2.033) & -2.031 (-2.203, -1.867) & -2.018 (-2.206, -1.821) \\
Age${}^{0.5}$ & & 30.603 (28.089, 33.030) & 28.381 (26.127, 30.639) & 29.063 (26.415, 31.577) \\
Ethnicity   & & 0.505 (0.278, 0.727) & 0.630 (0.391, 0.868) & 1.043 (0.747, 1.319) \\
Gender     & & -3.401 (-3.934, -2.888) & -3.733 (-4.219, -3.233) & -4.436 (-5.032, -3.827) \\
Marital Status & & 0.193 (0.033, 0.347) & 0.371 (0.222, 0.523) & 0.454 (0.270, 0.628) \\
 
\bottomrule

\end{tabular*}
\end{table}

\begin{figure}[t]
  \caption{Trace, density and autocorrelation plots for the risk factors of SBP at three quantile levels ($\tau=0.5,0.75,0.95$) under the Bayesian quantile regression model with FPs.}
  \includegraphics[width = 0.5\textwidth]{BQRwithFP_SBP_5.pdf}
  \includegraphics[width = 0.5\textwidth]{BQRwithFP_SBP_75.pdf}
  \includegraphics[width = 0.5\textwidth]{BQRwithFP_SBP_95.pdf}
  \label{fig:BQRwithFP-SBP}
\end{figure}

\begin{figure}[t]
  \caption{Trace, density and autocorrelation plots for the risk factors of DBP at three quantile levels ($\tau=0.5,0.75,0.95$) under the Bayesian quantile regression model with FPs.}
  \includegraphics[width = 0.5\textwidth]{BQRwithFP_DBP_5.pdf}
  \includegraphics[width = 0.5\textwidth]{BQRwithFP_DBP_75.pdf}
  \includegraphics[width = 0.5\textwidth]{BQRwithFP_DBP_95.pdf}
  \label{fig:BQRwithFP-DBP}
\end{figure}

\begin{figure}[t]
  \caption{Trace, density and autocorrelation plots for the risk factors of SBP at three quantile levels ($\tau=0.5,0.75,0.95$) under the Bayesian quantile regression model with FPs and variable selection.}
  \includegraphics[width = 0.5\textwidth]{BQRVSwithFP_SBP_5.pdf}
  \includegraphics[width = 0.5\textwidth]{BQRVSwithFP_SBP_75.pdf}
  \includegraphics[width = 0.5\textwidth]{BQRVSwithFP_SBP_95.pdf}
  \label{fig:BQRVSwithFP-SBP}
\end{figure}

\begin{figure}[t]
  \caption{Trace, density and autocorrelation plots for the risk factors of DBP at three quantile levels ($\tau=0.5,0.75,0.95$) under the Bayesian quantile regression model with FPs and variable selection.}
  \includegraphics[width = 0.5\textwidth]{BQRVSwithFP_DBP_5.pdf}
  \hfill
  \includegraphics[width = 0.5\textwidth]{BQRVSwithFP_DBP_75.pdf}
  \hfill
  \includegraphics[width = 0.5\textwidth]{BQRVSwithFP_DBP_95.pdf}
  \label{fig:BQRVSwithFP-DBP}
\end{figure}

\begin{figure}[t]
  \caption{The selected predictors and cutoff thresholds (dashed lines) of the NHANES dataset for the SBP model via the BQRVS-FP approach at $\tau = 0.50$, $\tau = 0.75$ and $\tau = 0.95$.}  
  \includegraphics[width = 0.5\textwidth]{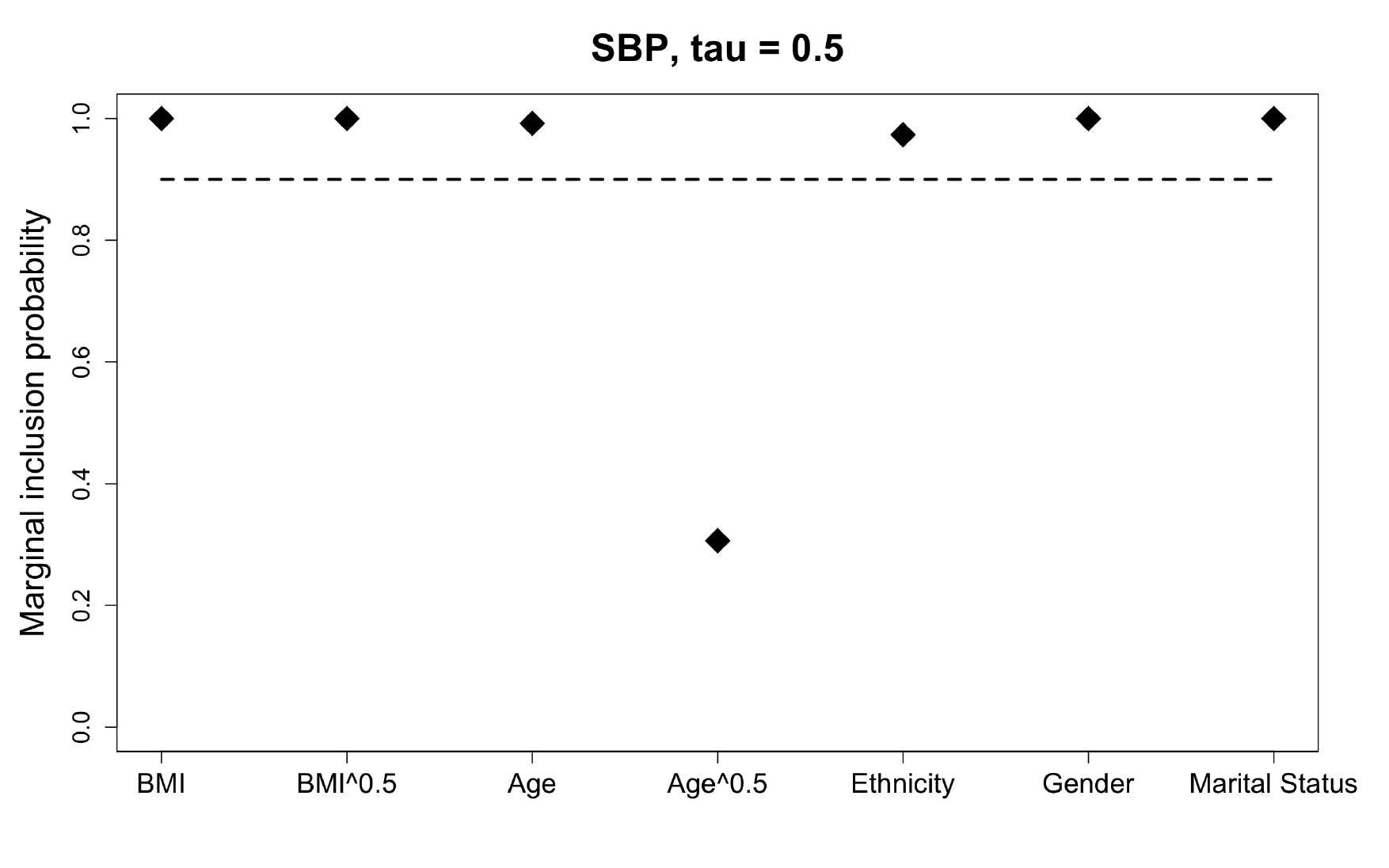}
  \includegraphics[width = 0.5\textwidth]{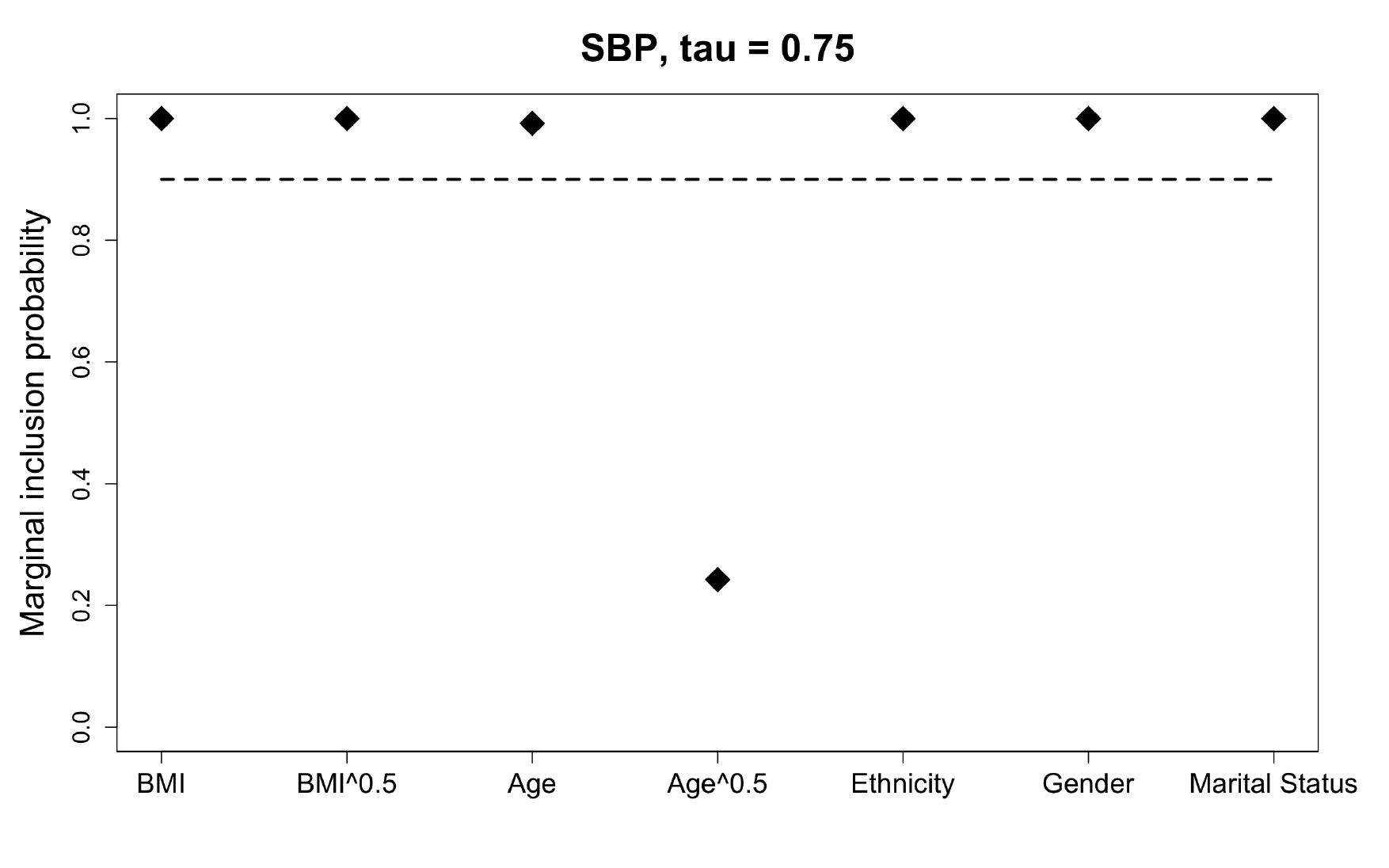}
  \includegraphics[width = 0.5\textwidth]{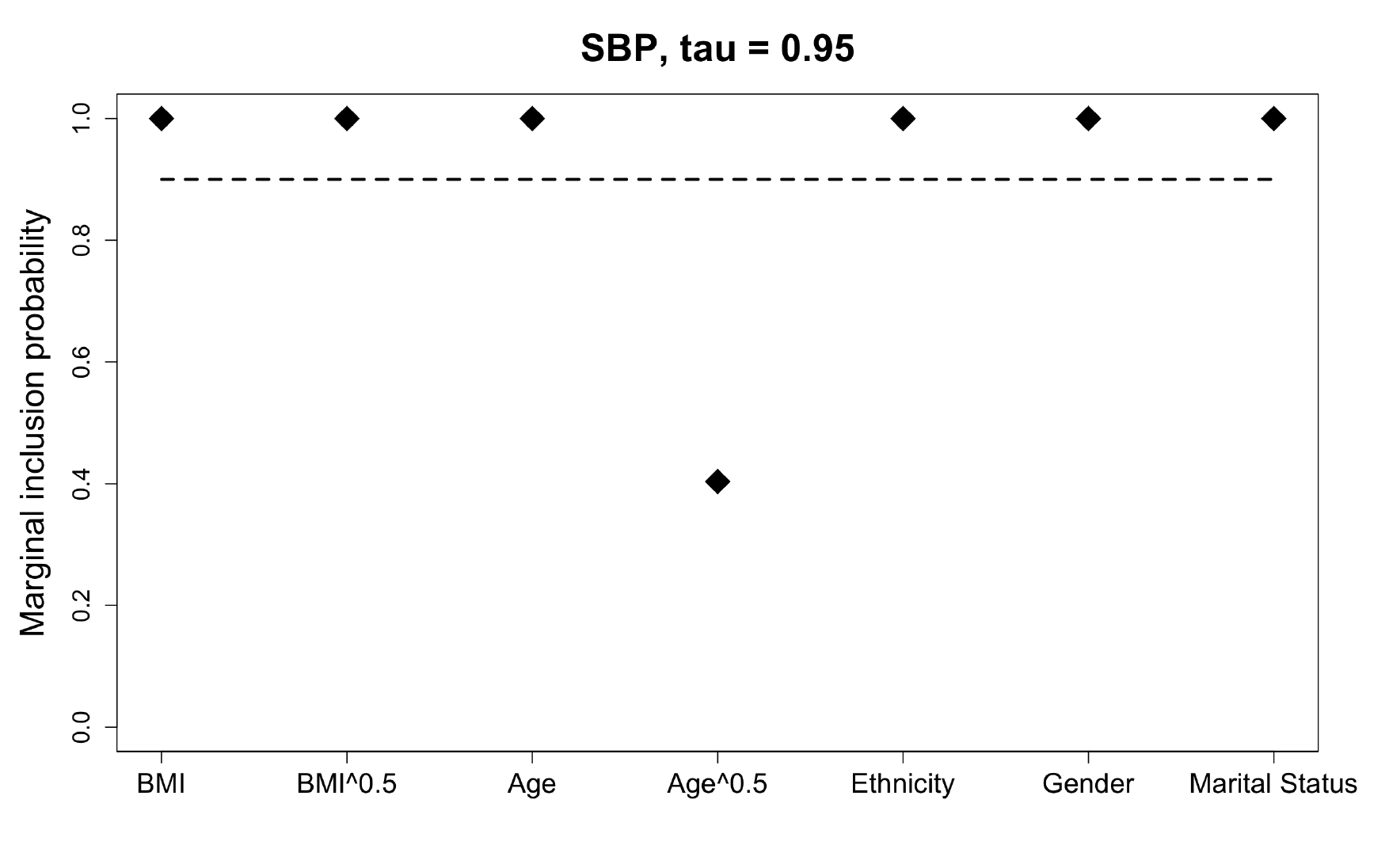} \label{fig:marginal-SBP}
\end{figure}

\begin{figure}[t]
   \caption{The selected predictors and cutoff thresholds (dashed lines) of the NHANES dataset for the DBP model via the BQRVS-FP approach at $\tau = 0.50$, $\tau = 0.75$ and $\tau = 0.95$.}
  \includegraphics[width = 0.5\textwidth]{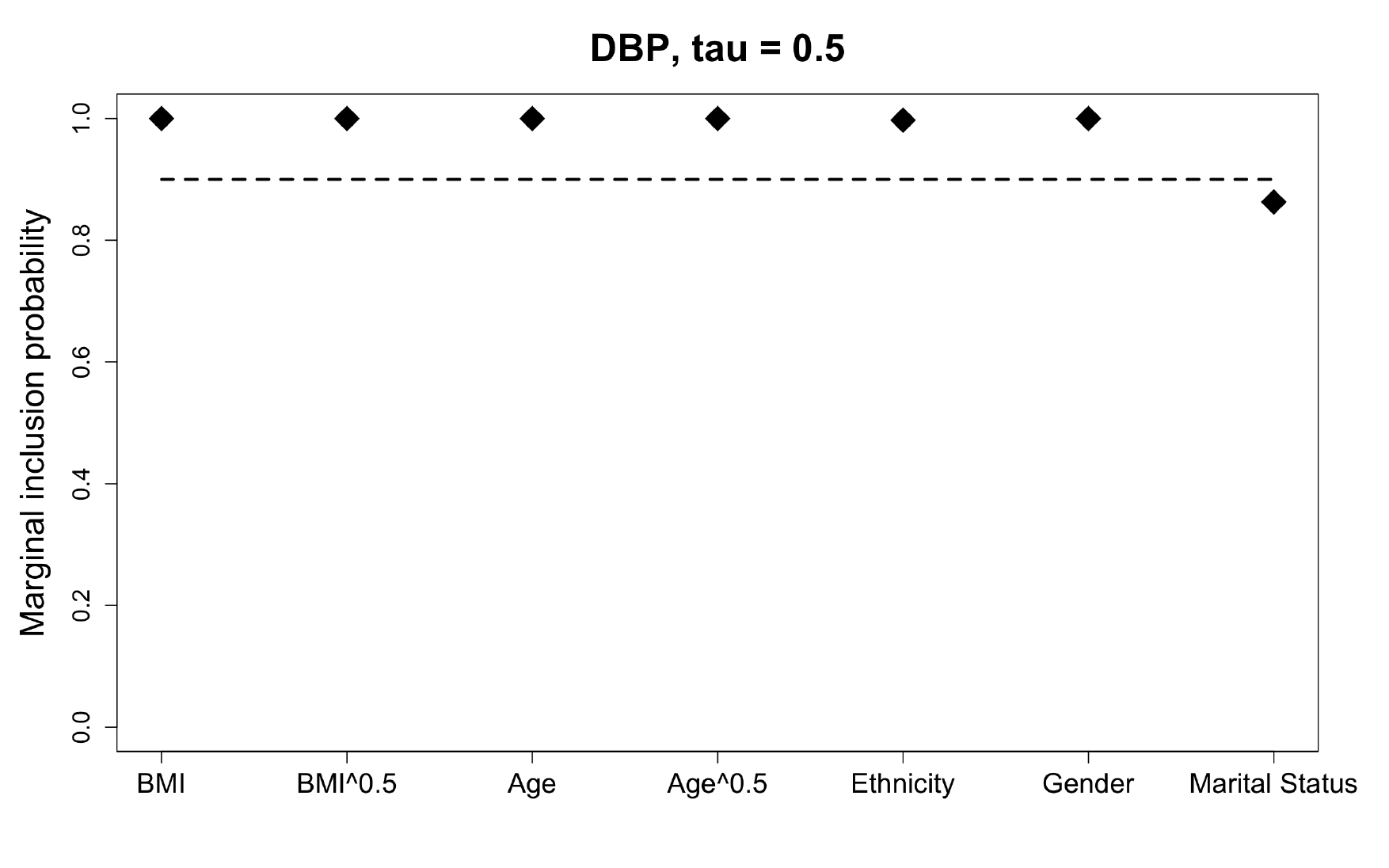}
  \includegraphics[width = 0.5\textwidth]{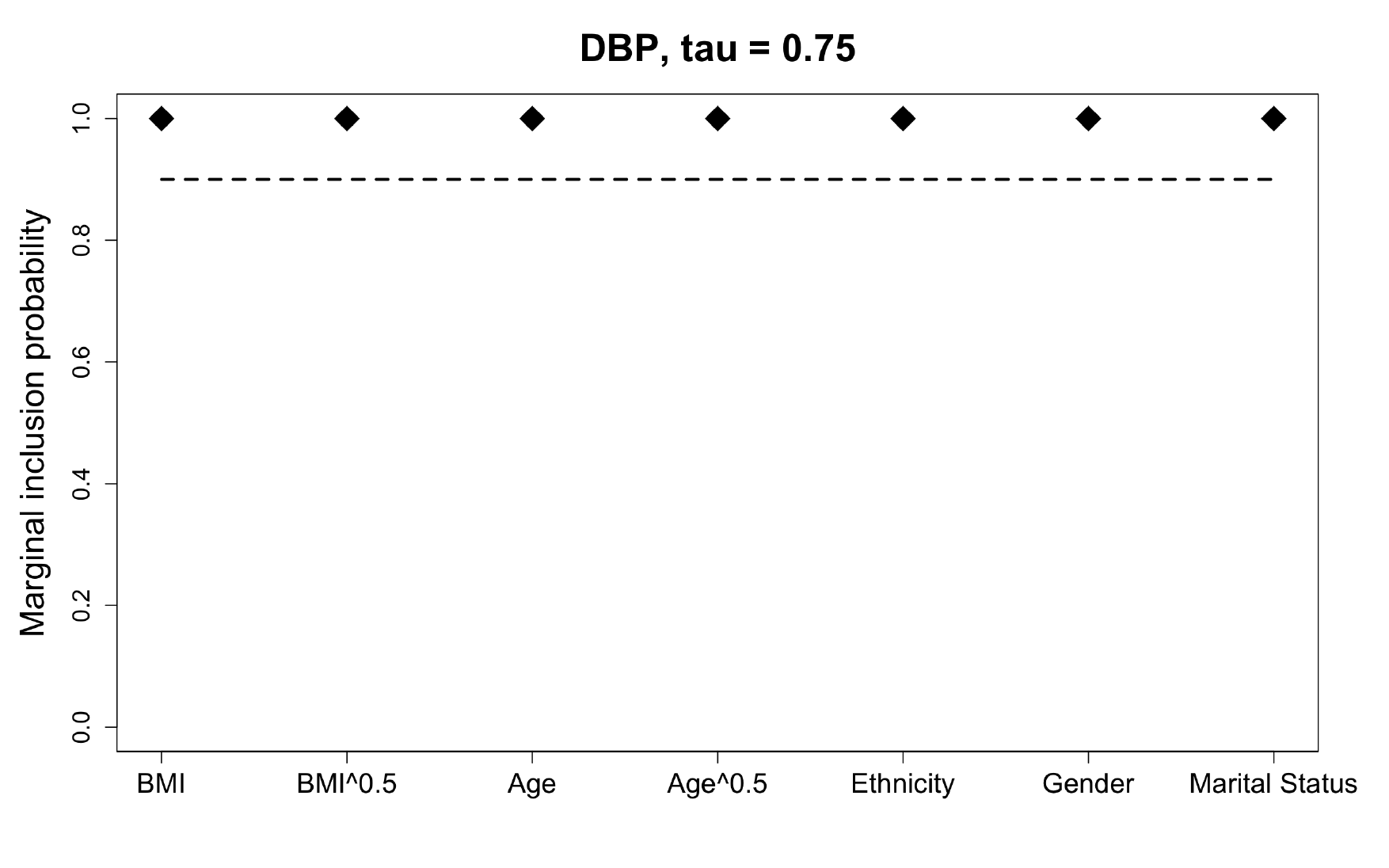}
  \includegraphics[width = 0.5\textwidth]{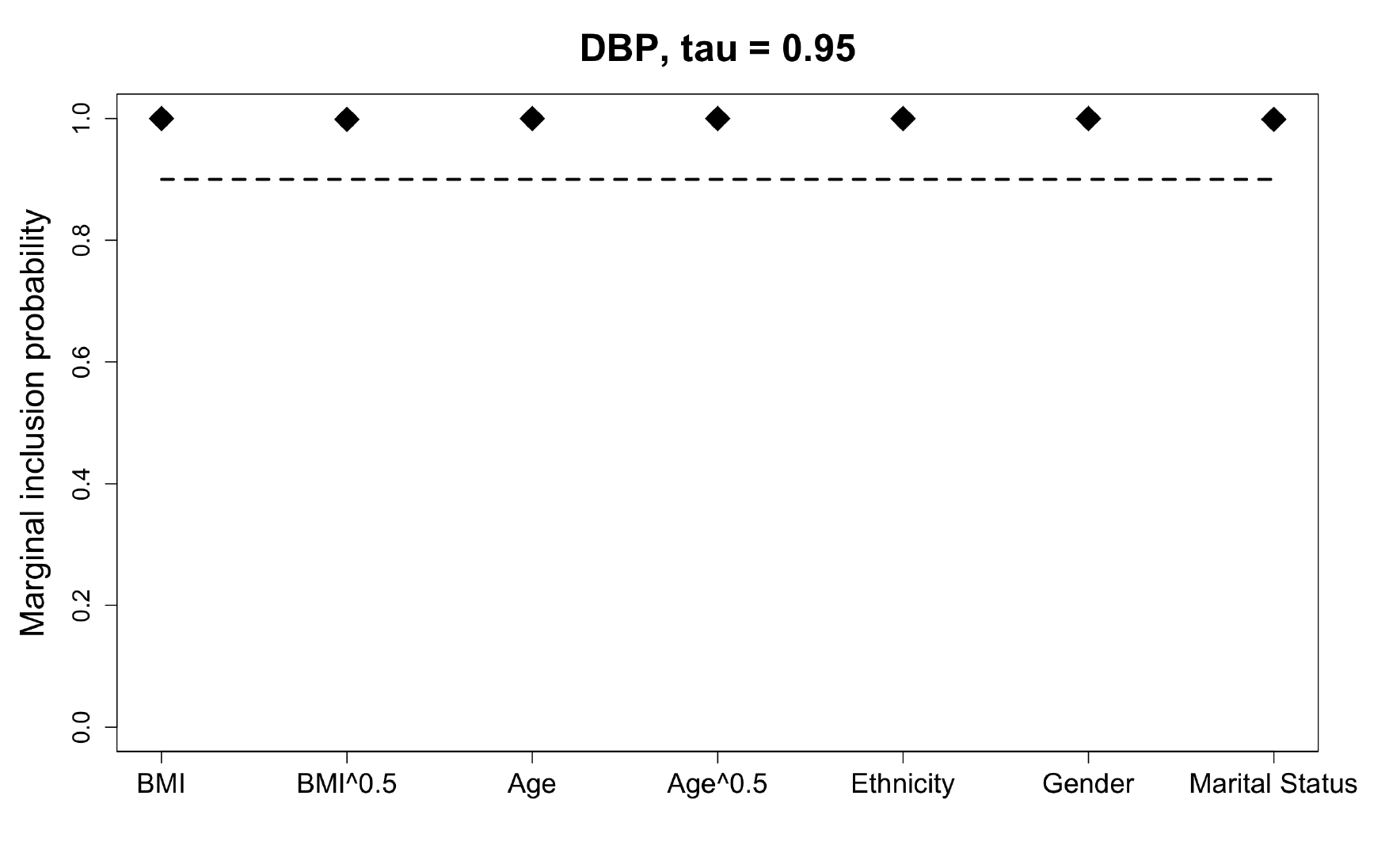}
  \label{fig:marginal-DBP}
\end{figure}

In this section, both descriptive and model analyses are provided for the NHANES 2007-2008 dataset using the proposed model. To evaluate the performance of the proposed model, we included two existing methods, including quantile regression and Bayesian quantile regression, with FP model for a fair comparative analysis. The model comparison is discussed outlining the advantages of the proposed model over these two methods. All the results are provided in this section through tables and figures for each regression analysis. 

\subsection{Descriptive Analysis}

For this analysis, continuous variables were collapsed into categorical variables, including SBP, DBP, BMI and age. According to the guidelines of \cite{WheltonEtAl2017}, the BP variables are divided into three groups: normal ($<120$ mmHg for SBP, $<80$ mmHg for DBP), pre-hypertension ($120-139$ mmHg for SBP, $80-89$ mmHg for DBP) and hypertension ($\geq 140$ mmHg for SBP, $\geq 90$ mmHg for DBP). The BMI variable is also divided into six groups: underweight ($<18.5$), healthy ($18.5-24.9$), overweight ($25-29.9$), obese ($30-34.9$), very obese ($35-39.9$) and morbidly obese ($\geq 40$) (\cite{CDC2022}). 

Table \ref{tab:SBPdes}-\ref{tab:DBPdes} present SBP and DBP proportions among US adults by demographic and lifestyle characteristics, including BMI, age, ethnicity, gender and marital status. 
The Cramer's V value is used to measure the magnitude of the association between SBP, DBP, sociodemographic characteristics and BMI of the participants. Their values with p-values are also presented in Table \ref{tab:SBPdes}-\ref{tab:DBPdes} and compared with with guidelines given by \cite{ReaParker2014}: 0.00 to under 0.10 = very weak association, 0.10 to under 0.20 = weak association, 0.20 to under 0.40 = moderate association and 0.40 and above = strong association. 

It is evident from Table \ref{tab:SBPdes}-\ref{tab:DBPdes} that hypertension was more prevalent in underweight, very obese and morbidly obese participants for both BP measures where the very obese and morbidly obese had the highest prevalence for DBP and SBP measures, respectively. The same trend was observed on the proportions of elevated BP for DBP measure. It was clear that healthy participants had the highest prevalence of normal BP for both BP measures. 

Concerning age, the prevalence of both elevated BP and hypertension increased with age, with the 40-49 years age group having the highest proportions for DBP measure and the 50 years and above age group for SBP measure. In regards to ethnicity, the non-Hispanic Black participants had the highest prevalence of hypertension compared to other race for both BP measures. 

Table \ref{tab:SBPdes}-\ref{tab:DBPdes} also showed that men had the highest prevalence of both elevated BP and hypertension for both BP measures. Participants who were separated or divorced and those who became widowed had the highest prevalence of hypertension for DBP and SBP measures, respectively. 

Lastly, at the $1\%$ significance level, Table \ref{tab:SBPdes}-\ref{tab:DBPdes} exhibited very weak to weak associations between BP measures, BMI and sociodemographic characteristics among the US adults. However, there is a moderate association between SBP measure and age. There is no statistically significant association between DBP measure and marital status at the $5\%$ level. 

\subsection{Model Analysis}

Table \ref{tab:SBPest}-\ref{tab:DBPest} provides the coefficients for predictors relating to SBP and DBP responses for three quantile regression models with FPs at three quantile levels ($\tau=0.50,0.75,0.95$), including one frequentist and two Bayesian approaches with one using variable selection. Bayesian parameters are obtained via posterior men. The 95\% confidence intervals are provided for the frequentist approach, whilst the 95\% credible intervals are provided for the Bayesian approaches. Either 95\% confidence interval or 95\% credible interval indicates that the user is 95\% confident that the population mean is within the interval. We denote the frequentist approach as the QR-FP model, and two Bayesian approaches as the BQR-FP and BQRVS-FP models where the latter uses variable selection. 

For the BQR-FP model, the algorithm was implemented for 10,000 MCMC iterations and 1,000 MCMC iterations were discarded as a burn-in period. For the BQRVS-FP model, the first stage algorithm ran for 1,000 EM iterations and repeated for 2 replications. Then 5,000 MCMC iterations were drawn for the second stage algorithm while discarding 2,500 MCMC iterations as a burn-in period. Finally, the last algorithm ran for 1,250 important re-weighting steps of which 500 steps were discarded as a burn-in period. The value of $g$ is selected as 1,000 for all implementations of the variable selection model. 

It is evident from Table \ref{tab:SBPest} that all the risk factors except both linear and nonlinear terms of age were found to have statistically significant associations with SBP across the two upper quantile levels according to their 95\% confidence intervals containing no zero value under the QR-FP model. Looking at the median level, the linear term had association with SBP under the same approach. When looking at the BQR-FP and BQRVS-FP models, only the nonlinear term of age did not have a statistically significant association for all quantile levels. On the other hand, Table \ref{tab:DBPest} observed that all the risk factors including nonlinear terms had statistically significant associations with DBP across all quantile levels for all model approaches. When looking at the median level under the QR-FP model, it revealed that the marital status did not have statistically significant association. 

Table \ref{tab:SBPest} also observed that the BMI, nonlinear term of age and gender have negative associations with SBP, whilst the nonlinear term of BMI, age and gender have negative associations with DBP from Table \ref{tab:DBPest} for all three model approaches. Under SBP model, the coefficients of BMI, ethnicity, gender and marital status increased when the quantile levels increased. The same trend was observed for the coefficients of BMI's nonlinear term, age, ethnicity and marital status under DBP model. Observing the coefficient of age's nonlinear term, all models saw the reverse U-shaped trend under SBP model and on other hand, both QR-FP and BQR-FP models had decreasing trend and the BQRVS-FP had U-shaped trend under DBP model. Increasingly, the coefficient of BMI's nonlinear term under SBP model followed the decreasing trend for the QR-FP model, the U-shaped trend for the BQR-FP model and the square-root trend for the BQRVS-FP model. 

Convergence of both Bayesian approaches was assessed using the trace plots, the density plots and autocorrelation plots. This is essential to perform various diagnostic tools for the assessing the convergence (\cite{Sinharay2003}). The convergence diagnostics are useful to check stationarity of the Markov chain or good chain mixing and to verify the accuracy of the posterior estimates (\cite{LesaffreLawson2012}). The trace plot is in form of a time series plot indicating whether it reaches stationarity or not. The density plot represents the stationary distribution of posterior samples approximating the posterior distribution of interest. The autocorrelation plot reports the correlation of posterior samples at each chain step with previous estimates that same variable, lagged by sample number of iterations. Decreasing trend indicates that the stationary distribution is more random and less dependent on initial values in the chain (\cite{HamraEtAl2013}).

Figure \ref{fig:BQRwithFP-SBP}-\ref{fig:BQRwithFP-DBP} present the trace, density and autocorrelation plots for each risk factor of SBP and DBP, respectively under the BQR-FP model. When looking at the trace plots across all the quantile levels, they exhibit stationarity due to relatively constant mean and variance of each plot. Thus, they show the good Markov chain mixing rate. When looking at the density plots across all the quantile levels, they reflect a smooth distribution with one peak at the mode of the distribution indicating a good convergence. It is also shown from Figure \ref{fig:BQRwithFP-SBP}-\ref{fig:BQRwithFP-DBP} that each risk factor of SBP and DBP across all the quantile levels has increasingly random stationary posterior distribution although at the $95^{th}$ percentile, the trend has a slower decreasing rate.

Figure \ref{fig:BQRVSwithFP-SBP}-\ref{fig:BQRVSwithFP-DBP} also present the trace, density and autocorrelation plots for each risk factor of SBP and DBP, respectively under the BQRVS-FP model. All the plots show stationarity, good Markov chain mixing rate and good convergence. Each autocorrelation plot indicated that their stationary distribution became random and less correlated with the initial values at a faster rate. 

Figure \ref{fig:marginal-SBP}-\ref{fig:marginal-DBP} provide the BQRVS-FP model determined by risk factors of SBP and DBP having the marginal inclusion probability (MIP) of at least 0.9, respectively. The risk factors selected lie above the cutoff of 0.9 of MIP. Across all the quantile levels for both SBP and DBP models, the same important risk factors were consistently selected where the DBP model selected all the risk factors including the nonlinear terms except marital status at the median level. The SBP model selected all except the nonlinear term of age. This mostly agreed with findings on $95\%$ credible intervals from Table \ref{tab:SBPest}-\ref{tab:DBPest}.

\subsection{Model Comparison}

Observing at the $95\%$ confidence intervals of frequentist approach and the 95\% credible intervals of two Bayesian approaches from Table \ref{tab:SBPest}-\ref{tab:DBPest}, the BQRVS-FP model has tighter intervals compared to the QR-FP model having wider intervals. 

Another finding is from the diagnostic plots that the autocorrelation plots of BQRVS-FP model have a faster decreasing trend rate across all the quantile levels, whereas those of the BQR-FP model have a slower rate. This is evident that the BQRVS-FP model has more random stationary posterior distributions of interest. 

When looking at Table \ref{tab:SBPest}-\ref{tab:DBPest} and Figure \ref{fig:marginal-SBP}-\ref{fig:marginal-DBP}, the BQRVS-FP model selected the important predictors coinciding with statistically significant associations between SBP, DBP and their risk factors based on their $95\%$ credible intervals.

These findings suggest that the Bayesian variable selection approach to quantile regression model with FPs obtained more precise estimates than the frequentist and Bayesian approaches. The nonlinear terms were selected as important variables in both SBP and DBP models indicating that FP model was necessary to examine the nonlinear relationship between SBP, DBP and risk factors. 

\section{Conclusion}\label{sec:conclusion}

In this paper, we conducted the data analysis of the impact of body mass index (BMI) on the blood pressure (BP) measures, including systolic and diastolic BP using data extracted from the 2007-2008 National Health and Nutrition Examination Survey (NHANES). The descriptive analysis showed that the prevalence of hypertension increases by age and the hypertension is highly prevalent among very obese and morbidly obese participants. In particular, it is more prevalent in men than women. Moreover, there is a statistically significant moderate association between SBP and age based on the Cramer's V value, whilst the remaining associations were weaker for both BP measures. However, there is no association between DBP and marital status. 

The analysis motivates a new Bayesian nonlinear quantile regression model under fractional polynomials (FPs) and variable selection with quantile-dependent prior where the quantile regression analysis investigates how the relationships differ across the median and upper quantile levels. The use of FPs allows for the relationships to be nonlinear parameterically. The variable selection investigates for important predictors that contribute to the nonlinear relationships via the Bayesian paradigm. The model analysis suggested that the proposed model provides better estimates because in comparison of two methods, the frequentist and Bayesian approaches of quantile regression model, the $95\%$ credible intervals were narrower and the autocorrelation plots have faster decreasing rate of correlated posterior samples. The analysis of the data showed that nonlinear relations do exist because the proposed model identified the nonlinear terms of continuous variables, including BMI and age as important predictors in the model across all the quantile levels. On the other hand, the nonlinear term of age is not selected under the SBP model. The marital status is not selected as an important risk factor for the DBP model at the median level. This agreed with findings of both descriptive and model analyses. Moreover, the data analysis suggested that the quantile based FP approaches have goodness of fit comparing to mean-based FP approaches. Thus, the importance of the nonlinear quantile model with FPs is significant for modelling of BP measures.

\section{Acknowledgments}
This work is supported by the UK Engineering and Physical Sciences Research Council (EPSRC) grant 2295266 for the Brunel University London for Doctoral Training.

\end{document}